\documentclass[preprint,aps]{revtex4}
\usepackage{graphicx}
\usepackage{epstopdf}
\usepackage{dcolumn}
\usepackage{bm}
\usepackage{color}

\begin{document}

\title{Electronic structure, imaging contrast and chemical reactivity of graphene moir\'e on metals}

\author{E. N. Voloshina,$^1$ E. Fertitta,$^1$ A. Garhofer,$^2$\\ F. Mittendorfer,$^2$ M. Fonin,$^3$ A. Thissen,$^4$ and Yu. S. Dedkov$^{4,}$\footnote{Corresponding author. E-mail: Yuriy.Dedkov@specs.com}}
\affiliation{$^1$Physikalische und Theoretische Chemie, Freie Universit\"at Berlin, 14195 Berlin, Germany}
\affiliation{$^2$Institute of Applied Physics, Vienna University of Technology, Gusshausstr. 25/134, 1040 Vienna, Austria}
\affiliation{$^3$Fachbereich Physik, Universit\"at Konstanz, 78457 Konstanz, Germany}
\affiliation{$^4$SPECS Surface Nano Analysis GmbH, Voltastra\ss e 5, 13355 Berlin, Germany}

\date{\today}

\begin{abstract}
\textbf{Realization of graphene moir\'e superstructures on the surface of $4d$ and $5d$ transition metals offers templates with periodically modulated electron density, which is responsible for a number of fascinating effects, including the formation of quantum dots and the site selective adsorption of organic molecules or metal clusters on graphene. Here, applying the combination of scanning probe microscopy/spectroscopy and the density functional theory calculations, we gain a profound insight into the electronic and topographic contributions to the imaging contrast of the epitaxial graphene/Ir(111) system. We show directly that in STM imaging the electronic contribution is prevailing compared to the topographic one. In the force microscopy and spectroscopy experiments we observe a variation of the interaction strength between the tip and high-symmetry places within the graphene moir\'e supercell, which determine the adsorption cites for molecules or metal clusters on graphene/Ir(111).}
\end{abstract}

\maketitle

Graphene layers on metal surfaces have been attracting the attention of scientists since several decades, starting from middle of the 60s, when the catalytic properties of the close-packed surfaces of transition metals were in the focus of the surface science research~\cite{Tontegode:1991ts,Wintterlin:2009,Batzill:2012,Dedkov:2012book}. The demonstration of the fascinating electronic properties of the free-standing graphene~\cite{Novoselov:2005,Zhang:2005}, renewed the interest in the graphene/metal systems, which are considered as the main and the most perspective way for the large-scale preparation of high-quality graphene layers with controllable properties~\cite{Yu:2008,Kim:2009a,Li:2009,Bae:2010}. For this purpose single-crystalline as well as polycrystalline substrates of $3d-5d$ metals can be used. 

One of the particularly exciting questions concerning the graphene/metal interface is the origin of the bonding mechanism in such systems~\cite{Wintterlin:2009,Batzill:2012,Dedkov:2012book,Voloshina:2012c}. This graphene-metal puzzle is valid for both cases: graphene adsorption on metallic surfaces as well as for the opposite situation of the metal deposition on the free-standing or substrate-supported graphene. In the latter case the close-packed surfaces of $4d$ and $5d$ metals are often used as substrates~\cite{NDiaye:2009a,Sicot:2012}. A graphene layer prepared on such surfaces, i.\,e. Ru(0001)~\cite{Marchini:2007,Wang:2010jw,Altenburg:2010ke}, Rh(111)~\cite{Wang:2010ky,Sicot:2012,Voloshina:2012a}, Ir(111)~\cite{Coraux:2009,Busse:2011}, or Pt(111)~\cite{Land:1992,Sutter:2009a}, forms so-called moir\'e structures due to the relatively large lattice mismatch between graphene and metal substrates. As a consequence of the lattice mismatch the interaction strength between graphene and the metallic substrate is spatially modulated leading to the spatially periodic electronic structure. Such lateral graphene superlattices are known to exhibit selective absorption for organic molecules~\cite{Zhang:2012} or metal clusters~\cite{Sicot:2010}. Especially, the adsorption of different metals - Ir, Ru, Au, or Pt - on graphene/Ir(111) has been intensively studied showing a preferential nucleation around the so-called $FCC$ or $HCP$ high-symmetry positions within the moir\'e unit cell~\cite{NDiaye:2009a,Knudsen:2012ei}. In the subsequent works~\cite{Feibelman:2008,Knudsen:2012ei} this site-selective adsorption was explained via local $sp^2$ to $sp^3$ rehybridization of carbon atoms with the bond formation between graphene and the cluster. However, a fully consistent description of the local electronic structure of graphene/Ir(111), the observed imaging contrast in scanning probe experiments and the bonding mechanism of molecules or clusters on it is still lacking, motivating the present research.

Here we present the systematic studies of the graphene/Ir(111) system by means of density functional theory (DFT) calculations and scanning tunnelling and atomic force microscopy (STM and AFM) performed in constant current / constant frequency shift (CC / CFS) and constant height (CH) modes. The obtained results for the graphene/Ir(111) system allow to separate the topographic and electronic contributions in the imaging contrast in STM and AFM as well as to shed light on the spatially modulated interaction between graphene/Ir(111) and the metallic STM/AFM tip, which is of paramount importance for the understanding of absorption of metals on top of graphene/Ir(111) as well as of similar graphene-metal systems.


\section*{Results}

\begin{figure}
\centering
\includegraphics[scale=2.2]{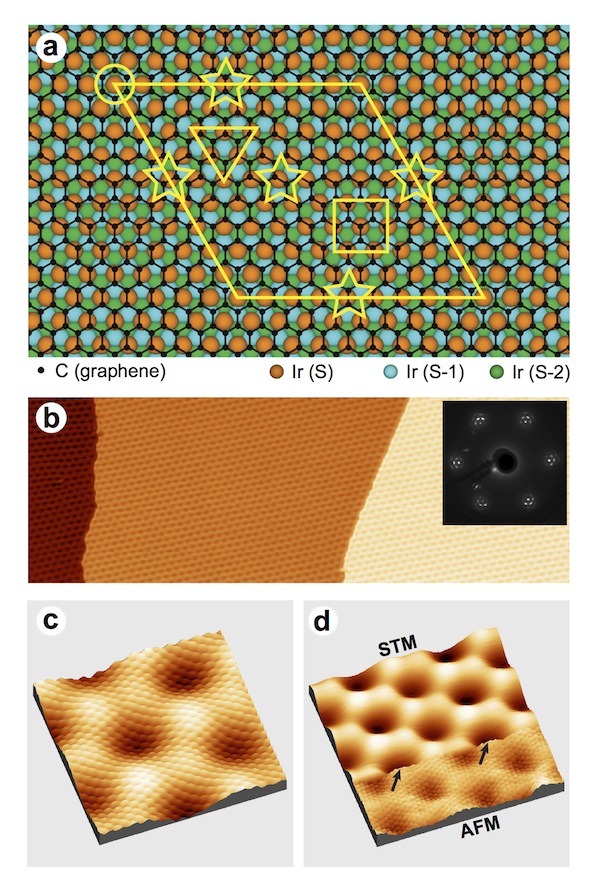}\\
\caption{\label{model-stm} (a) Crystallographic structure of $(10\times10)$ graphene on $(9\times9)$ Ir(111). The high-symmetry places are marked by circle, rectangle, down-triangle, and stars for $ATOP$, $FCC$, $HCP$, and $BRIDGE$ positions. (b) Large scale ($162\times55$\,nm$^2$) STM image of graphene/Ir(111). The inset shows the corresponding LEED image obtained at 71\,eV. (c) Atomically resolved STM image of graphene/Ir(111) showing the inverted contrast. (d) A combined STM/AFM image with a switching between CC STM and CFS AFM imaging during scanning. The scan sizes are $5.2\times5.2$\,nm$^2$ ($U_T=+550$\,meV, $I_T=540$\,pA) and $10\times10$\,nm$^2$ (STM: $U_T= +461$\,meV, $I_T=7.6$\,nA, AFM: $\Delta f=-675$\,mHz), respectively.} 
\end{figure}

The unit cell of graphene on Ir(111) is shown in Fig.~\ref{model-stm}(a) with the corresponding high-symmetry local arrangements of carbon atoms above Ir layers marked in the figure: $ATOP$ (circles), $FCC$ (squares), $HCP$ (down-triangles), and $BRIDGE$ (stars). The DFT-D2 optimized local distances between graphene and Ir(111) are $3.27$\,\AA\ ($HCP$), $3.28$\,\AA\ ($FCC$), $3.315$\,\AA\ ($BRIDGE$), $3.58$\,\AA\ ($ATOP$). This result is very close to the recently published equilibrium structure for this system~\cite{Busse:2011}. The similar distances for $HCP$, $FCC$, and $BRIDGE$ positions can be related to the fact that in all these cases one of the carbon atoms in the graphene unit cell is placed above Ir(S) atom defining the local interaction strength for the particular high-symmetry position. The obtained distances are very close to those between carbon layers in pure graphite and this was explained by a binding interaction dominated by van der Waals effects between graphene and Ir(111) that is modulated by weak bonding interactions at the $FCC$ and $HCP$ places and anti-bonding chemical interaction around $ATOP$ positions~\cite{Busse:2011}. As a result a small charge transfer from the graphene $\pi$ states on Ir empty valence band states is detected in calculations and the Dirac point is shifted by $\approx 100$\,meV above the Fermi level ($E_F$) that is close to the data obtained by photoelectron spectroscopy~\cite{Pletikosic:2009} [see Fig.~\ref{biases}(c) and discussion below].

The graphene/Ir(111) system is a nice example of the moir\'e structure, which is easily recognisable in LEED and on the large scale STM images shown in Fig.~\ref{model-stm}(b). The extracted lattice parameter of this structure from LEED and STM is $25.5$\,\AA\ and $25.2$\,\AA, respectively, that is in good agreement with previously published data~\cite{NDiaye:2008qq}. In most cases, for the typical bias voltages used in STM imaging, the graphene/Ir(111) structure is imaged in the so-called \textit{inverted contrast} [Fig.~\ref{model-stm}(c,d)]~\cite{NDiaye:2008qq}, when topographically highest $ATOP$ places are imaged as dark and topographically lowest $FCC$ and $HCP$ as bright regions. This assignment was initially done in Ref.~\cite{NDiaye:2008qq}, where the STM topography of neighbouring regions of the clean and graphene-covered Ir(111) surface were imaged. In our studies we perform the ``on-the-fly'' switching between CC STM and CFS AFM imaging during scanning [Fig.~\ref{model-stm}(d)] where we observe the inversion of the topographic contrast $z(x,y)$ (see also Fig.\,S1 of the supplementary material~\cite{supl}). This becomes clearly evident around areas marked by arrows in Fig.~\ref{model-stm}(d) where the darkest contrast in CC STM becomes the brightest one in CFS AFM for the $ATOP$ position. In the latter case the darkest areas correspond to $HCP$ sites that correlates with the calculated height changes between lowest and highest carbon positions. The pronounced difference between the $FCC$ and $HCP$ areas can be explained by fact that additional bias voltage was applied in order to increase the atomically-resolved contrast.

The inversion of the imaging contrast was also detected in CC STM images when the bias voltage is changed from $-0.5$\,V to $-1.8$\,V during scanning [Fig.~\ref{biases}(a)]. The images taken at positive voltages are similar to the lower part of the figure Fig.~\ref{biases}(a) (see Fig.\,S2 of the supplementary material~\cite{supl}). The simulations of STM images using the experimental conditions show very good agreement with the obtained data [Fig.~\ref{biases}(a,b) and Fig.\,S2]. The presented STM results demonstrate the big difference in the local electronic structure for high-symmetry positions of carbon atoms on Ir(111). That can points out on the difference in the local adsorption strength for these places.

\begin{figure}
\centering
\includegraphics[scale=2]{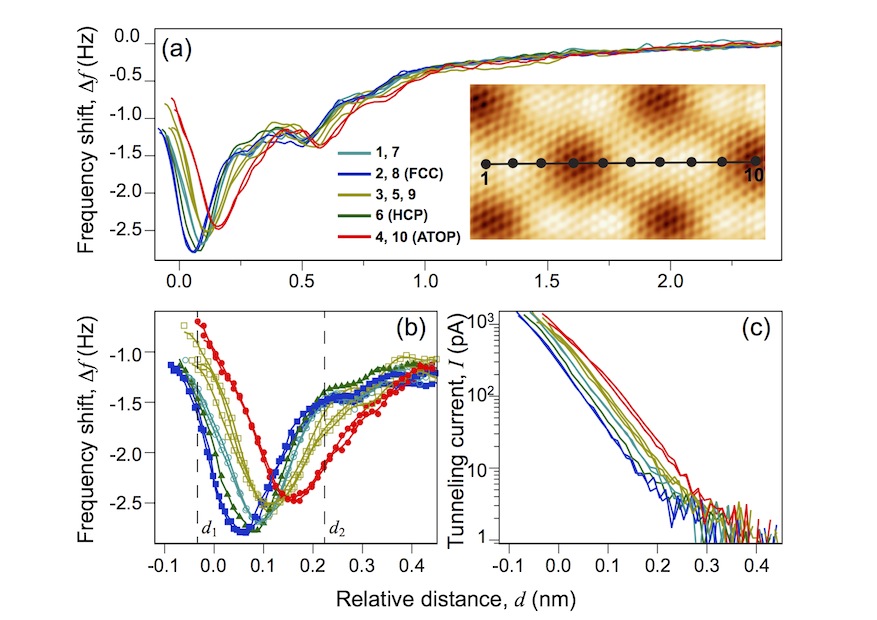}\\
\caption{\label{freqcur} Frequency shift (a,b) of the oscillating KolibriSensor and the corresponding tunnelling current (c) as a function of the relative distance $d$ between tip and the graphene/Ir(111) sample. These curves were acquired with respect to the set-point for the CC STM imaging. Inset of (a) shows the corresponding STM image ($U_T=+50$\,mV, $I_T=400$\,pA) with the path where $\Delta f$ and $I$ data were measured. }
\end{figure}

In order to prove this assertion we have performed the force microscopy and spectroscopy experiments on graphene/Ir(111). Fig.~\ref{freqcur} shows (a,b) the frequency shift of an oscillating scanning sensor as a function of a distance from the sample, $\Delta f(d)$, and (c) the corresponding tunnelling current, $I(d)$, measured in the unit cell of graphene/Ir(111) along the path marked in the CC STM image shown as an inset of (a). During these measurements the tunnelling current was used for the stabilisation of the feedback loop, allowing to determine the relative $z$-position of the $\Delta f$ and $I$ curves. Following the sequence: $F_z(d)=-\partial E(d)/\partial d$, $\Delta f(d)=-f_0/2k_0\cdot \partial F_z(d)/\partial d$, where $E(d)$ and $F_z(d)$ are the interaction energy and the vertical force between a tip and the sample, respectively, $f_0$ and $k_0$ are the resonance frequency and the spring constant of the sensor, correspondingly, we can separate repulsive, attractive, and long-range electrostatic and van der Waals contributions in AFM imaging. The imaging contrast at distances more than $5$\,\AA\ from the surface is defined by long-range van der Waals interactions which are insensitive to the atomically resolved structure of the scanning tip and the sample. The short-range chemical interactions around the minimum of the $\Delta f$ curve are dominated either by repulsive (left-hand side) or attractive (right-hand side) forces and give the atomically-resolved site-selective chemical contrast.

The presented $\Delta f(d)$ curves [Fig.~\ref{freqcur}(a,b)] clearly show the site-selective interaction in the unit cell of the graphene/Ir(111) system. First of all, the absolute value of the maximal frequency shift is higher for the $FCC$ and $HCP$ areas compared to $ATOP$ ones indicating the stronger interaction of graphene with the tungsten STM/AFM tip for the former carbon positions. We would like to note, that the same trend will be observed for the adsorption of metals on graphene/Ir(111). Secondly, the difference in the positions of $\Delta f(d)$-curve minima is $0.097$\,nm [Fig.~\ref{freqcur}(b)] compared to the value of graphene corrugation of $0.067$\,nm obtained from the CC STM image [inset of Fig.~\ref{freqcur}(a)], also pointing on the difference in the interaction strength for two carbon positions, $ATOP$ and $FCC$.

The measured $\Delta f(d)$ curves were used to track and explain the contrast obtained in CH AFM experiments (Fig.~\ref{constheight-d_f-I} and Fig.\,4 in Supplementary material~\cite{supl}). In this experiments the $z$-coordinate of the oscillating sensor was fixed during scanning and $\Delta f(x,y)$ and $I(x,y)$ maps were collected. As can be clearly seen from the presented results the imaging contrast in CH $\Delta f$ images is changing as a function of $z$-coordinate and it is fully inverted for two limit $z$-positions [$d_1=-0.033$\,nm and $d_2=+0.227$\,nm, see Fig.~\ref{freqcur}(b)] used for the imaging (upper row in Fig.~\ref{constheight-d_f-I}). In the region of distances between the tip and the graphene/Ir(111) sample where chemical forces determine the interaction, on the right-hand side of the $\Delta f$ curve the interaction is attractive giving the more negative frequency shift for the $ATOP$ positions as located closer to the oscillating tip compared to the $FCC$ and $HCP$ areas. On the other side, on the left-hand side of the curve the repulsive interaction starts to contribute in the interaction giving the smaller frequency shift for $ATOP$ compared to other positions that can again be explained by the different distances between tip and the sample for different high-symmetry positions as well as by the different interaction strength at these places. At the same time the contrast for the current map $I(x,y)$ is always the same, \textit{inverted contrast} (lower row in Fig.~\ref{constheight-d_f-I}). Taking into account that during CH imaging at $+50$\,meV in Fig.~\ref{constheight-d_f-I} the imaging contrast for $I(x,y)$ is \textit{inverted} due to the higher DOS for the $FCC$ and $HCP$ positions and that variation of the distance for CH imaging and CC STM imaging (when $U_T$ is changed from $-0.5$\,V to $-1.8$\,V) is nearly the same, one can separate the topographic and electronic contributions into imaging at different biases and distances.

\begin{figure}
\centering
\includegraphics[scale=2.25]{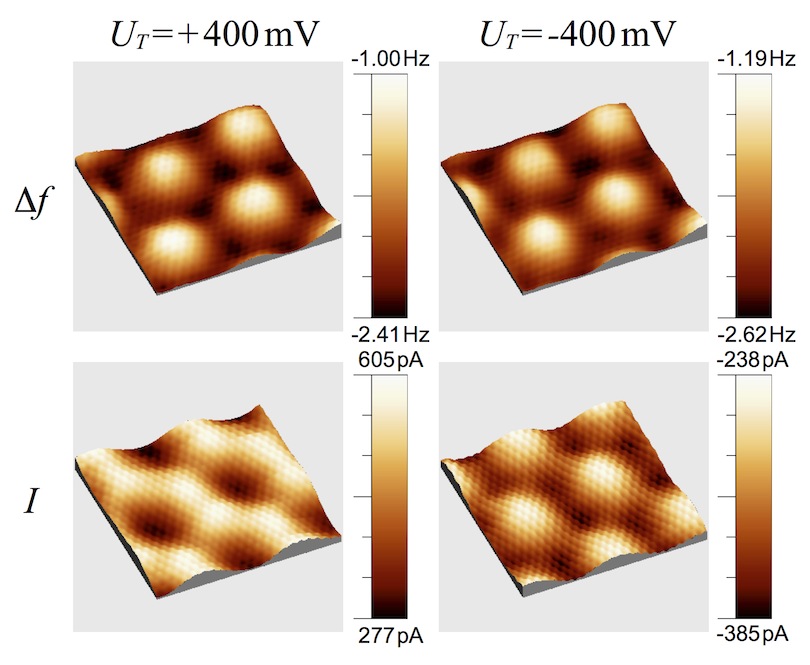}\\
\caption{\label{constheight-atres} Atomically-resolved constant height images of graphene/Ir(111), $\Delta f(x,y)$ (upper row) and $I(x,y)$ (lower row), collected at opposite signs bias voltages. The size of the scanning area is $4.3\times4.3$\,nm$^2$.}
\end{figure}

Here we would also like to note that the change of the bias voltage during CH AFM imaging does not lead to any changes in the imaging  contrast for $\Delta f$ (Fig.~\ref{constheight-atres}). The first row shows the two small-scanning range atomically resolved $\Delta f(x,y)$ images acquired on the same place of the graphene/Ir(111) sample with opposite signs for the bias voltage: the imaging contrast is the same with the slight variation of the imaging scale that can be explained by the small drift of the oscillating tip. The first look on the $I(x,y)$ map (lower row) might give an impression that the contrast is fully inverted. However, the absolute value of the tunnelling current is the same and only more negative values of the tunnelling current are shown as darker areas in the image for current.

\section*{Discussion}

\begin{figure}
\centering
\includegraphics[scale=1.25]{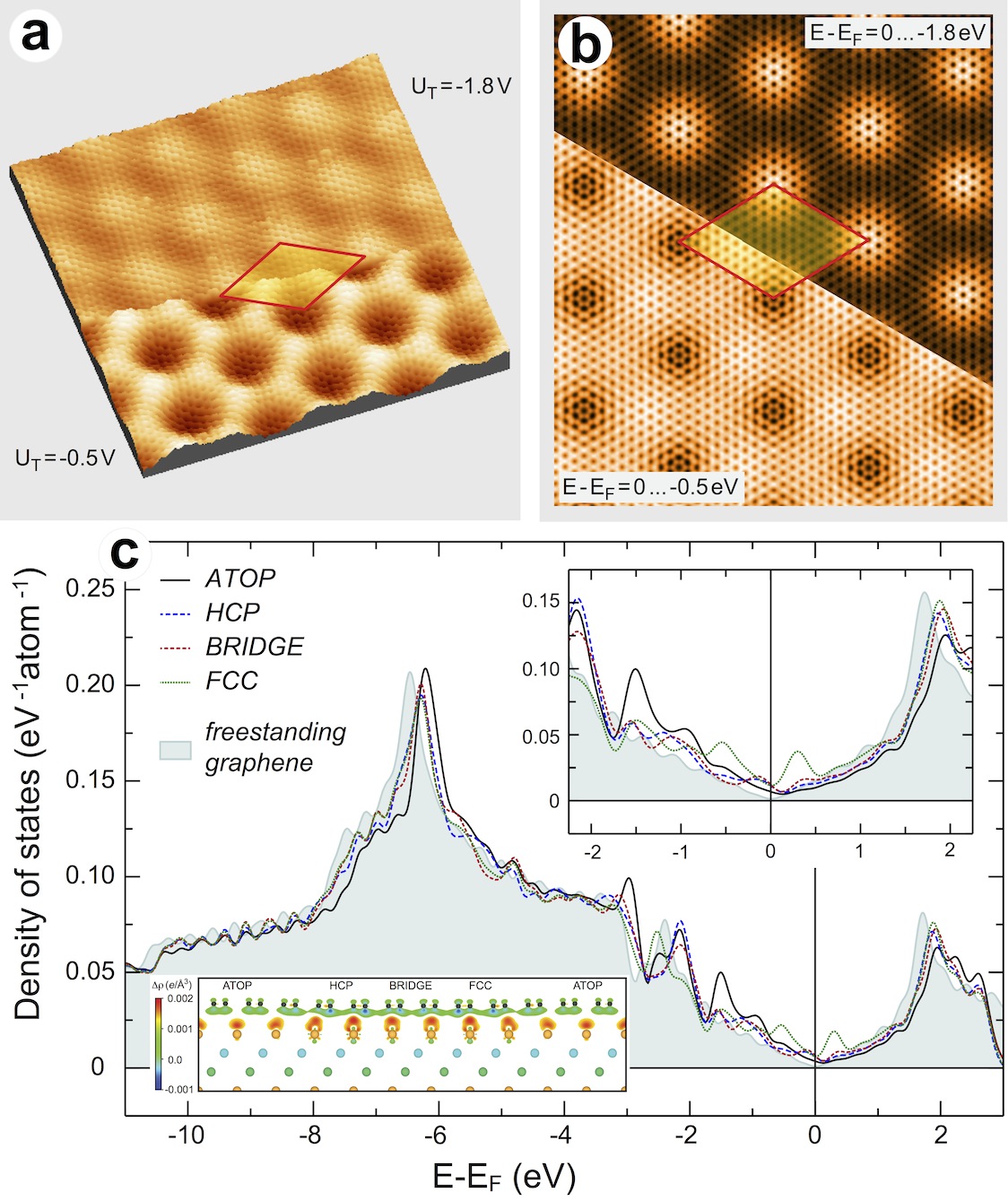}\\
\caption{\label{biases} Experimental (a) and calculated (b) CC STM images of graphene/Ir(111) obtained at $-0.5$\,V (bottom) and $-1.8$\,V (top) of the bias voltage ($I_T=15$\,nA). (c) Carbon-site projected density of states calculated for all high-symmetry places of graphene/Ir(111) (zoom around $E_F$ is shown as an upper inset). Difference electron density, $\Delta \rho(r)=\rho_{gr/Ir(111)}(r)-\rho_{Ir}(r)-\rho_{gr}(r)$, plotted in units of $e/$\AA$^3$ is shown as an inset.}
\end{figure}

The graphene/Ir(111) system was studied with DFT methods in oder to deeply understand the effects of the electronic structure on the adsorption properties of this system. Here we present the comparison between experimental and theoretical STM data. The STM images are calculated using the Tersoff-Hamann formalism~\cite{Tersoff:1985}, in its most basic formulation, approximating the STM tip by an infinitely small point source~\cite{Vanpoucke:2008,Voloshina:2011NJP,Voloshina:2012a}. In these simulations the constant current condition was fulfilled that leads to the increasing of the distance between the sample and a tip from $2.50$\,\AA\ for $U_T=-0.5$\,V to $3.21$\,\AA\ for $U_T=-1.8$\,V [Fig.~\ref{biases}(b)]. Taking into account that at negative bias voltages the electrons tunnel from occupied states of the sample into unoccupied states of the tip and that tunnelling current is proportional to the surface local density of states (LDOS) at the position of the tip we can clearly identify the states in the valence band of graphene/Ir(111) which are responsible for the formation of the imaging contrast (see Fig.\,S3 of the supplementary material for the identification of the local valence band states states). [The surface Brillouin zone (BZ) of the graphene/Ir(111) system is approximately 10 times smaller compared to BZ of free-standing graphene giving the maximal wave-vector of electrons of $k_{||}\approx0.17$\,\AA$^{-1}$ for the $K$-point, i.\,e. in this case one has to consider a small region in the reciprocal space around the $\Gamma$-point. In this case we can assign the features in the tunnelling current as originating from the corresponding peaks in DOS of the whole system (in general, the tunnelling current depends on the $k_{||}$ of electronic states participating in the tunnelling process and the electronic states with a large parallel component of the wave vector give a small contribution to the current)]. Therefore, the so-called inverted contrast for the small bias voltages, negative or positive, can be explained by two features in LDOS for the carbon atoms in the $FCC$ position located at $E-E_F=-0.53$\,eV and $E-E_F=+0.32$\,eV, respectively [Fig.~\ref{biases}(c)]. The next two maxima in LDOS for C $ATOP$ positions located at $E-E_F=-0.95$\,eV and $E-E_F=-1.49$\,eV [Fig.~\ref{biases}(c)] explain the fact that these places become brighter for higher bias voltages in experimental and simulated CC STM images [Fig.~\ref{biases}(a,b)].

The calculated LDOS for graphene/Ir(111) compared with DOS for the free-standing graphene [($10\times10$) unit cell] is shown in Fig.~\ref{biases}(c) together with the electron density difference distribution presented as an inset. The obtained DOS results demonstrate the slight difference in the position of main features ($\pi$-states at $E-E_F\approx6.1$\,eV, $\sigma$-states at $E-E_F\approx3.2$\,eV, and the $M$-point-derived Kohn anomaly at $E-E_F\approx2.2$\,eV) as well as the Dirac cone between two systems, that is assigned to the $p$-doping of graphene on Ir(111). The DOS features responsible for the formation of the STM contrast can be considered as a result of the hybridisation between graphene\,$\pi$ states and the valence band states of Ir around the particular high-symmetry adsorption positions (see Fig.\,S3 of the supplementary material for the identification of the local valence band states states). Here we would like to note, that one can also consider the increased LDOS around $E_F$ for $FCC$, $HCP$, and $BRIDGE$ positions as a hint indicating that these places can be a nucleation centers for adsorbed atoms.

\begin{figure}
\centering
\includegraphics[scale=1.85]{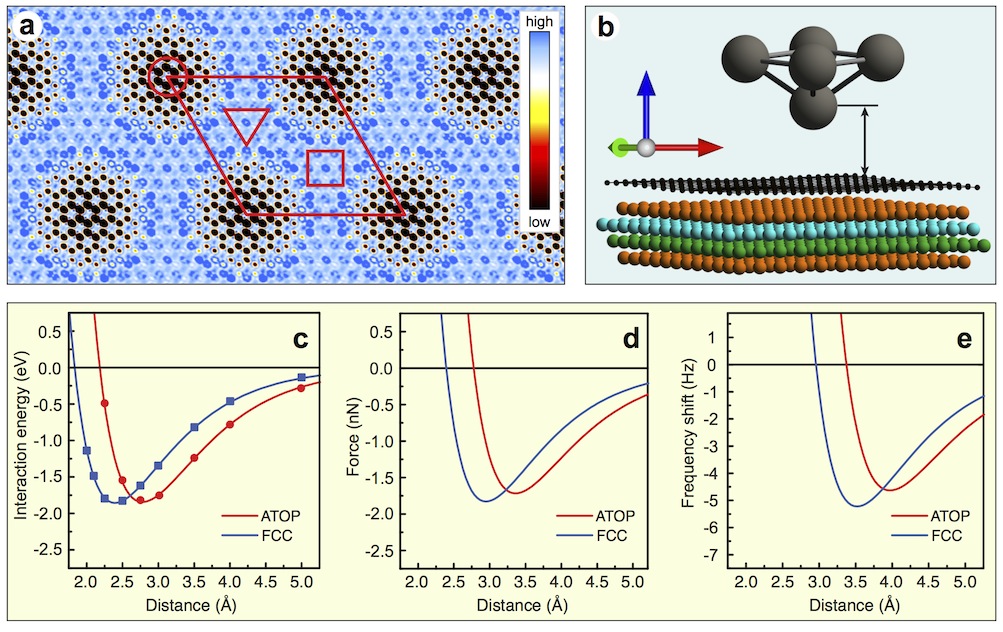}\\
\caption{\label{afm_theory} (a) Simulated CH AFM image of the graphene/Ir(111) system according to the approach suggested in Ref.~\cite{Chan:2009jb}. (b) Schematic representation of the geometry of the 5-atom-W-tip/graphene/Ir(111) system used in simulation of the frequency shift curves. Interaction energy (c) and the force (d) between 5-atom W-tip and graphene/Ir(111) calculated for two limit places $ATOP$ and $FCC$. (e) Frequency shift as a function of distance between the model 5-atom W-tip and graphene/Ir(111).}
\end{figure}

In oder to get better insight in the results obtained during AFM experiments we performed simulations of these data in the framework of DFT formalism (we would like to note that a qualitative description of the buckled graphene systems was performed in Ref.~\cite{Castanie:2012bv}, where the model Lennard-Jones potential was used to model tip-sample interaction). Due to the large size of the unit cell of the graphene/Ir(111) system we selected two approaches. In the first one, which requires less computational resources, the tip-sample force is expressed as a function of the potential $V_{ts}(\mathbf{r})$ on the tip due to the sample: \mbox{$F_{ts}\propto - \nabla [| \nabla V_{ts}(\mathbf{r})|^2]$}~\cite{Chan:2009jb}. However, this method does not take into account the geometrical and electronic structure of the scanning tip and thereby does not allow to get absolute values for the force or frequency shift and the correct distance between the tip and the sample, and only qualitative result for the attractive region of the interaction can be obtained. The result of such simulations for graphene/Ir(111) is shown in Fig.~\ref{afm_theory}(a), which is in rather good agreement with experimental data: the regions around $ATOP$ positions are imaged as dark compared to other high-symmetry cites imaged as brighter contrast. Unfortunately, the information about repulsive region of the forces can not be obtained from such calculations. However, if the model corrugation-dependent repulsive potential is added, then the resulting simulated $\Delta f$ curve reproduces qualitatively the experimentally obtained results.

For the purpose to reproduce our data in a more quantitative way, the interaction between the tip and the graphene/Ir(111) system was simulated within the second approach, where W-tip is approximated by the 5-atom pyramid as shown in Fig.~\ref{afm_theory}(b). The results of these calculations are compiled in Fig.~\ref{afm_theory}(c-e). The interaction energy (system was rigid without allowing to relax) between model W-tip and the surface was calculated for two limiting places of graphene/Ir(111), $ATOP$ and $FCC$. The calculated points are shown by filled rectangles and circles in Fig.~\ref{afm_theory}(c) for the $FCC$ and $ATOP$ positions, respectively. The Morse potential was used to fit the calculated data as the most suitable for the graphene-metal systems~\cite{Rafii:2003,Loske:2009}. The resulting curves are shown by solid lines in the same figure. The force and the frequency shift curves were calculated on the basis of the obtained interaction energy curves according to formulas presented earlier. The results are shown in Fig.~\ref{afm_theory}\,(d) and (e), respectively.

The interaction energy curves [Fig.~\ref{afm_theory}(c)] are very similar for the two high-symmetry areas and shifted with respect to each other by $0.40$\,\AA\ reflecting the height difference between $ATOP$ and $FCC$ regions. The maximal absolute value of energy is slightly higher for $FCC$ by $0.015$\,eV. However, our calculations do not take into account the relaxation of the system during the tip-sample interaction. Therefore the actual values can be different. Calculation of the force and the frequency shift via derivation of the respective curves leads to the clear discrimination in the value of the extrema as well as to the further separation of their positions and for the $\Delta f$ curves the extrema are separated by $0.45$\,\AA. The resulting values for the frequency shift for our model W-tip are approximately 2 times larger compared to experimental values. But, the shape and the trend for the inversion of the imaging contrast are clearly reproducible in our calculated data. The following reasons can explain the difference between experiment and theory: (i) small W-cluster modelling the actual W-tip, (ii) system was not relaxed during calculations, (iii) the actual atomic structure of the tip during experiment is unknown, (iv) small uncertainty in the tip position during measurements of $\Delta f$ curves. However, in spite of these simplifications, the presented calculation results allow us to confirm and better understand the data obtained during the AFM measurements.

\textit{In conclusion}, we performed the systematic investigations of the geometry, electronic structure and their effects on the observed imaging contrast during STM and AFM experiments on graphene/Ir(111). Our results obtained via combination of DFT calculations and scanning probe microscopy imaging allow to discriminate the topographic and electronic contributions in these measurements and to explain the observed contrast features. We found that in STM imaging the electronic contribution is prevailing compared to the topographic one and the inversion of the contrast can be assigned to the particular features in the electronic structure of graphene on Ir(111). Contrast changes observed in constant height AFM measurements are analyzed on the basis of the energy, force, and frequency shift curves reflecting the interaction of the W-tip with the surface and are attributed to the difference in the height and the different interaction strength for high-symmetry cites within the moir\'e unit cell of graphene on Ir(111). The presented findings are of general importance for the understanding of the properties of the lattice-mismatched graphene/metal systems especially with regard to possible applications as a template for molecules or clusters.

\section*{Methods}

\textit{DFT calculations.} The crystallographic model of graphene/Ir(111) presented in Fig.~\ref{model-stm}(a) was used in DFT calculations, which were carried out using the projector augmented plane wave method~\cite{Blochl:1994}, a plane wave basis set with a maximum kinetic energy of 400\,eV and the PBE exchange-correlation potential~\cite{Perdew:1996}, as implemented in the VASP program~\cite{Kresse:1994}. The long-range van der Waals interactions were accounted for by means of a semiempirical DFT-D2 approach proposed by Grimme~\cite{Grimme:2006}. The studied system is modelled using supercell, which has a $(9\times9)$ lateral periodicity and contains one layer of $(10\times10)$ graphene on a four-layer slab of metal atoms. Metallic slab replicas are separated by ca.~20\,\AA\ in the surface normal direction. To avoid interactions between periodic images of the slab, a dipole correction is applied~\cite{Neugebauer:1992}. The surface Brillouin zone is sampled with a $(3\times3\times1)$ $k$-point mesh centered the $\Gamma$ point.

\textit{STM and AFM experiments.} The STM and AFM measurements were performed in two different modes: constant current (CC) or constant frequency shift (CFS) and constant height (CH). In the first case the topography of sample, $z(x,y)$, is studied with the corresponding signal, tunnelling current ($I_T$) or frequency shift ($\Delta f$), used as an input for the feedback loop. In the later case the $z$-coordinate of the scanning tip is fixed with a feedback loop switched off that leads to the variation of the distance $d$ between tip and sample. In such experiments $I_T(x,y)$ and $\Delta f(x,y)$ are measured for different $z$-coordinates. The STM/AFM images were collected with Aarhus SPM 150 equipped with KolibriSensor\texttrademark\ from SPECS~\cite{Torbruegge:2010cf,Voloshina:2012a} with Nanonis Control system. In all measurements the sharp W-tip was used which was cleaned \textit{in situ} via Ar$^+$-sputtering. In presented STM images the tunnelling bias voltage, $U_T$, is referenced to the sample and the tunnelling current, $I_T$, is collected by the tip, which is virtually grounded. During the AFM measurements the sensor was oscillating with the resonance frequency of $f_0=999161$\,Hz and the quality factor of $Q=45249$. The oscillation amplitude was set to $A = 100$\,pm or $A = 300$\,pm.

\textit{Preparation of graphene/Ir(111).} The graphene/Ir(111) system was prepared in ultra-high vacuum station for STM/AFM studies according to the recipe described in details in Ref.~\cite{NDiaye:2008qq} via cracking of ethylene: $T=1100^\circ$\,C, $p=5\times10^{-8}$\,mbar, $t=5$\,min. This procedure leads to the single-domain graphene layer on Ir(111) of very high quality that was verified by means of low-energy electron diffraction (LEED) and STM. The base vacuum was better than $8\times10^{-11}$\,mbar during all experiments. All measurements were performed at room temperature.

\section*{Acknowledgements} 

The computing facilities (ZEDAT) of the Freie Universit\"at Berlin and the High Performance Computing
Network of Northern Germany (HLRN) are acknowledged for computer time. This work has been supported by the European Science Foundation (ESF) under the EUROCORES Programme EuroGRAPHENE (Project ``SpinGraph''). E.V. appreciates the support from the German Research Foundation (DFG) through the Collaborative Research Center (SFB) 765 ``Multivalency as chemical organization and action principle: New architectures, functions and applications''. M.F. gratefully acknowledges the financial support by the Research Center ``UltraQuantum'' (Excellence Initiative).







\newpage

\begin{figure}
\centering
\includegraphics[scale=2.25]{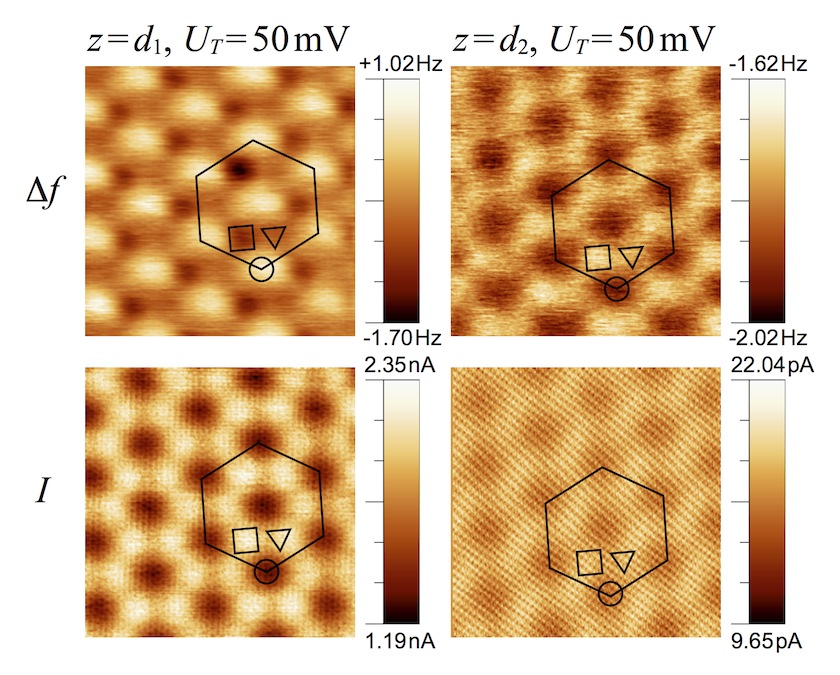}\\
\caption{\label{constheight-d_f-I} Constant height images of graphene/Ir(111), $\Delta f(x,y)$ (upper row) and $I(x,y)$ (lower row), obtained at two different heights $d_1$ and $d_2$ of the KolibriSensor above the surface (see Fig.~\ref{freqcur}). The size of the scanning area is $10.5\times10.5$\,nm$^2$.}
\end{figure}

\newpage
\noindent
Supplementary material for manuscript:\\
\textbf{Electronic structure, imaging contrast and chemical reactivity of graphene moir\'e on metals}\\
\newline
E. N. Voloshina,$^1$ E. Fertitta,$^1$ A. Garhofer,$^2$ F. Mittendorfer,$^2$ M. Fonin,$^3$ A. Thissen,$^4$ and Yu. S. Dedkov$^{4}$\\
\newline
$^1$Physikalische und Theoretische Chemie, Freie Universit\"at Berlin, 14195 Berlin, Germany\\
$^2$Institute of Applied Physics, Vienna University of Technology, Gusshausstr. 25/134, 1040 Vienna, Austria\\
$^3$Fachbereich Physik, Universit\"at Konstanz, 78457 Konstanz, Germany\\
$^4$SPECS Surface Nano Analysis GmbH, Voltastra\ss e 5, 13355 Berlin, Germany
\newline
\newline
\textbf{List of figures:}
\\
\newline
\noindent\textbf{Fig.\,S1.} STM/AFM images of the graphene/Ir(111) system obtained via switching between CC STM and CFS AFM modes ``on-the-fly''. Scanning parameters: (a) STM: $U_T=+461$\,meV, $I_T=7.6$\,nA, AFM: $\Delta f=-675$\,mHz (attractive regime), (b) STM: $U_T=-201$\,meV, $I_T=7.6$\,nA, AFM: $\Delta f=+675$\,mHz (repulsive regime).
\newline
\noindent\textbf{Fig.\,S2.} (a) STM image of graphene/Ir(111) obtained at $U_T=+0.7$\,V and $I_T=15$\,nA. (b) Calculated STM image obtained in the framework of the Tersoff-Hamann formalism via integration of the valence band states in the range $E-E_F=0...+0.5$\,eV. The corresponding distance between tip and the sample surface is $2.2$\,\AA.
\newline
\noindent\textbf{Fig.\,S3.} Band structures of graphene/Ir(111) calculated for the corresponding arrangements of the expanded graphene layer on Ir(111) in $(1\times1)$ structure. The distance between graphene and Ir(111) is fixed to the corresponding spacing in the graphene/Ir(111) nanomesh. The thickness of the energy bands corresponds to the higher atom- and state-projected contribution to the corresponding band.
\newline
\noindent\textbf{Fig.\,S4.} The sequence of the CH AFM images ($\Delta f$ and $I$) obtained at different $z$-positions of the scanning tip and correspondingly at different distances $d$ between tip and the surface. The bias voltage applied to the oscillating tip during scanning is $+50$\,meV.

\clearpage
\begin{figure}
\includegraphics[width=\textwidth]{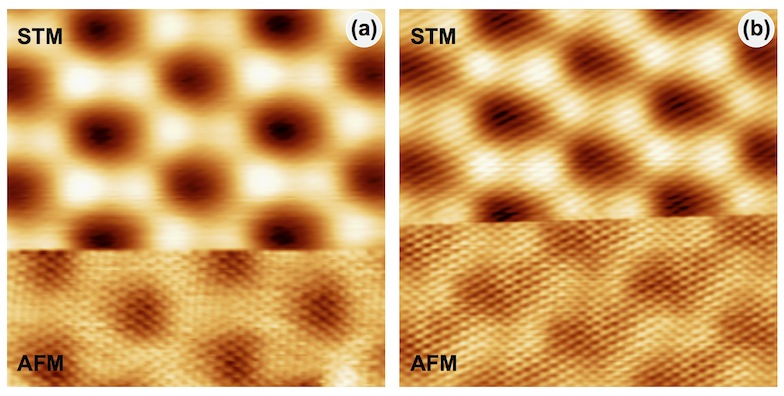}
\end{figure}
\noindent\textbf{Fig.\,S1.} STM/AFM images of the graphene/Ir(111) system obtained via switching between CC STM and CFS AFM modes ``on-the-fly''. Scanning parameters: (a) STM: $U_T=+461$\,meV, $I_T=7.6$\,nA, AFM: $\Delta f=-675$\,mHz (attractive regime), (b) STM: $U_T=-201$\,meV, $I_T=7.6$\,nA, AFM: $\Delta f=+675$\,mHz (repulsive regime).

\clearpage
\begin{figure}
\includegraphics[width=\textwidth]{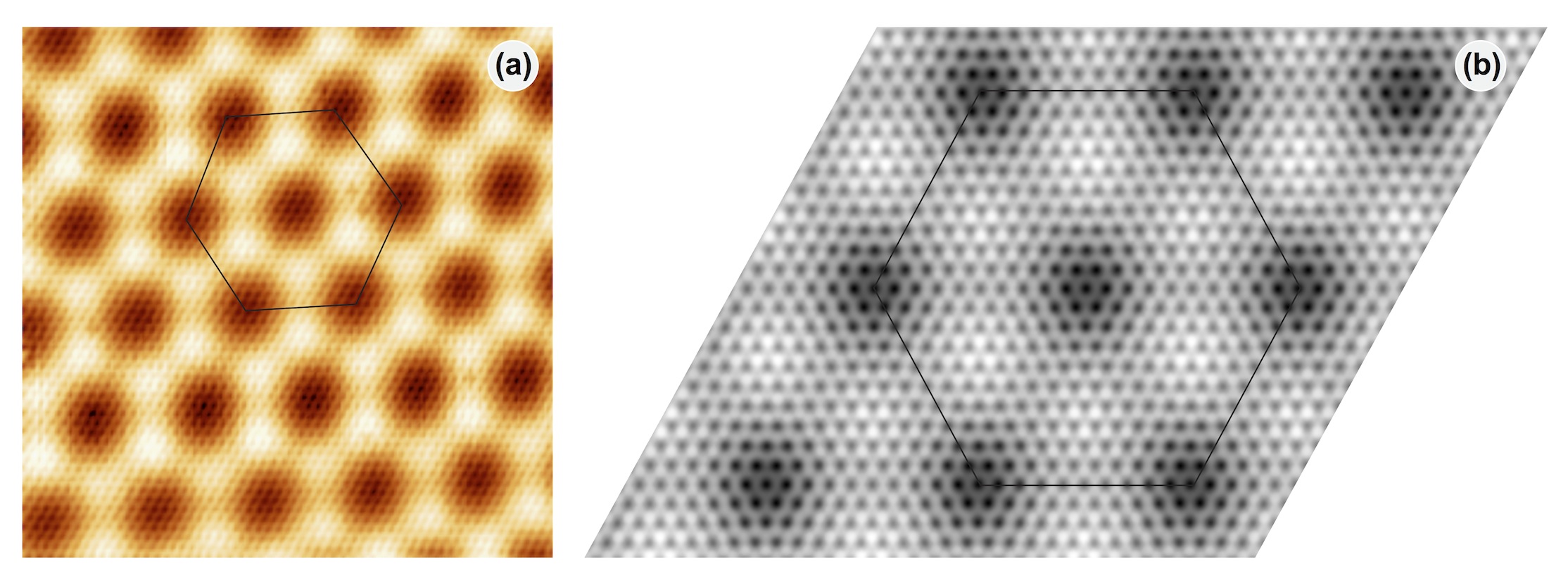}
\end{figure}
\noindent\textbf{Fig.\,S2.} (a) STM image of graphene/Ir(111) obtained at $U_T=+0.7$\,V and $I_T=15$\,nA. (b) Calculated STM image obtained in the framework of the Tersoff-Hamann formalism via integration of the valence band states in the range $E-E_F=0...+0.5$\,eV. The corresponding distance between tip and the sample surface is $2.2$\,\AA.

\clearpage
\begin{figure}
\includegraphics[width=\textwidth]{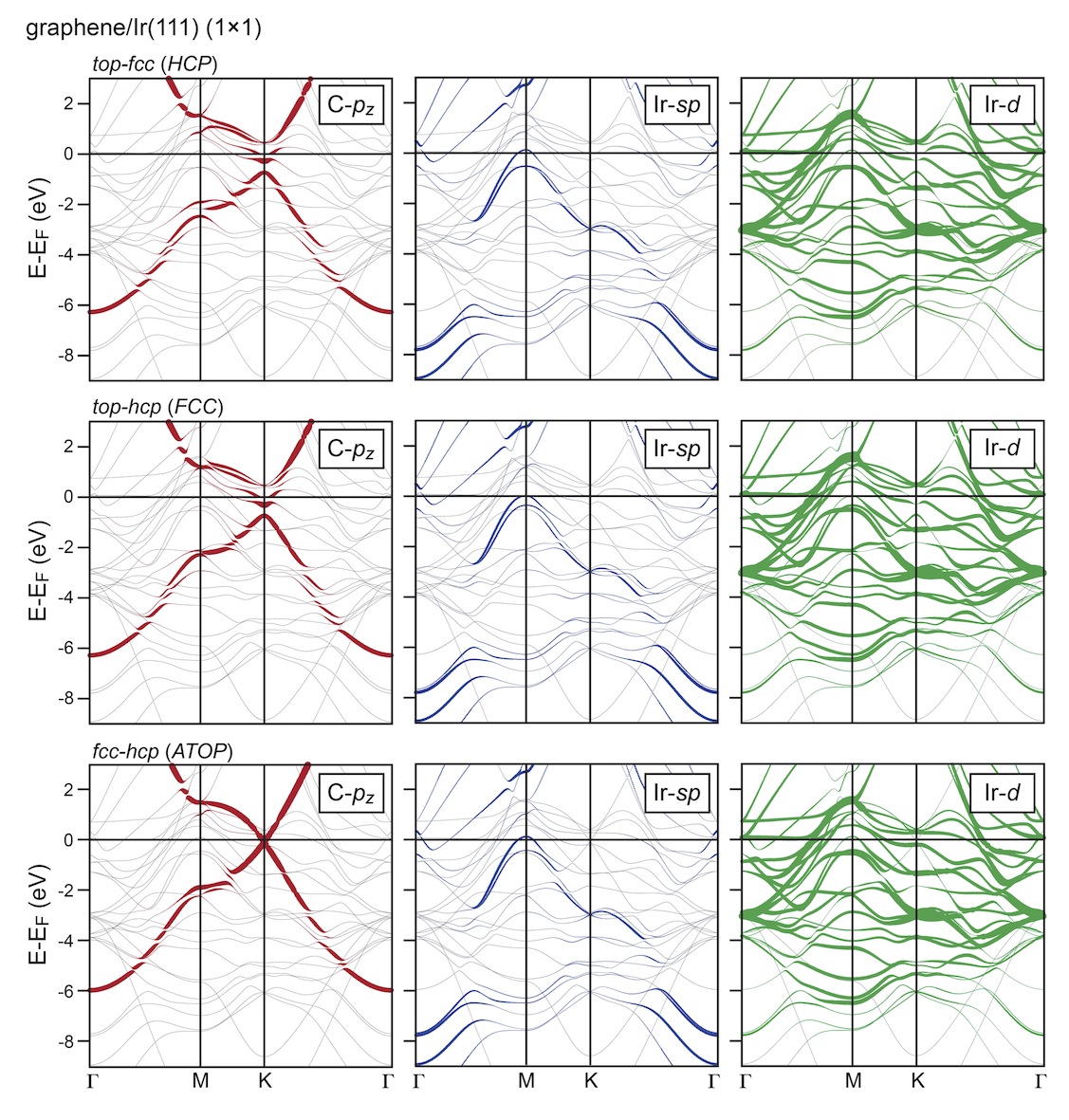}
\end{figure}
\noindent\textbf{Fig.\,S3.} Band structures of graphene/Ir(111) calculated for the corresponding arrangements of the expanded graphene layer on Ir(111) in $(1\times1)$ structure. The distance between graphene and Ir(111) is fixed to the corresponding spacing in the graphene/Ir(111) nanomesh. The thickness of the energy bands corresponds to the higher atom- and state-projected contribution to the corresponding band. 

\clearpage
\begin{figure}
\includegraphics[width=1.1\textwidth]{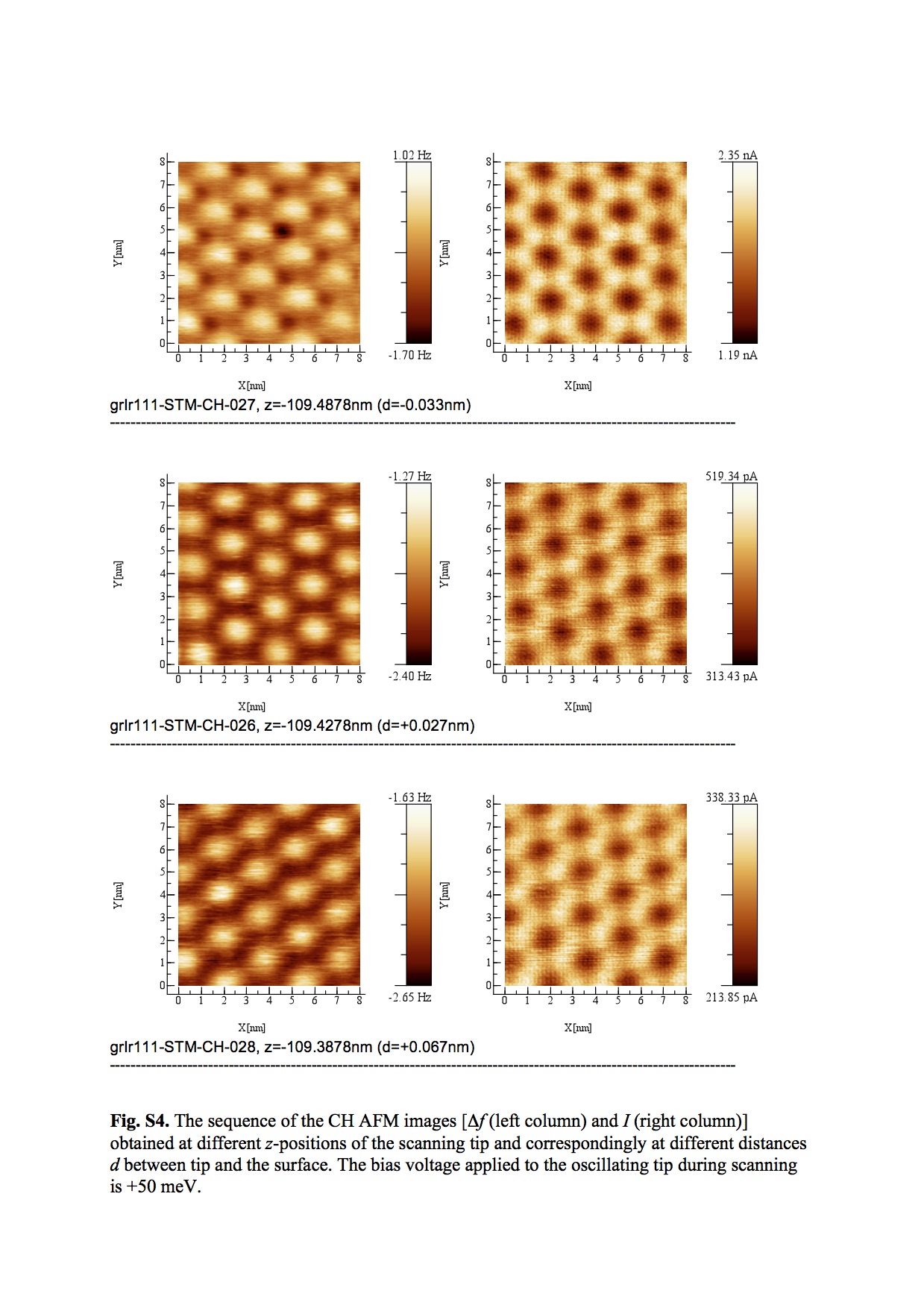}
\end{figure}

\clearpage
\begin{figure}
\includegraphics[width=1.1\textwidth]{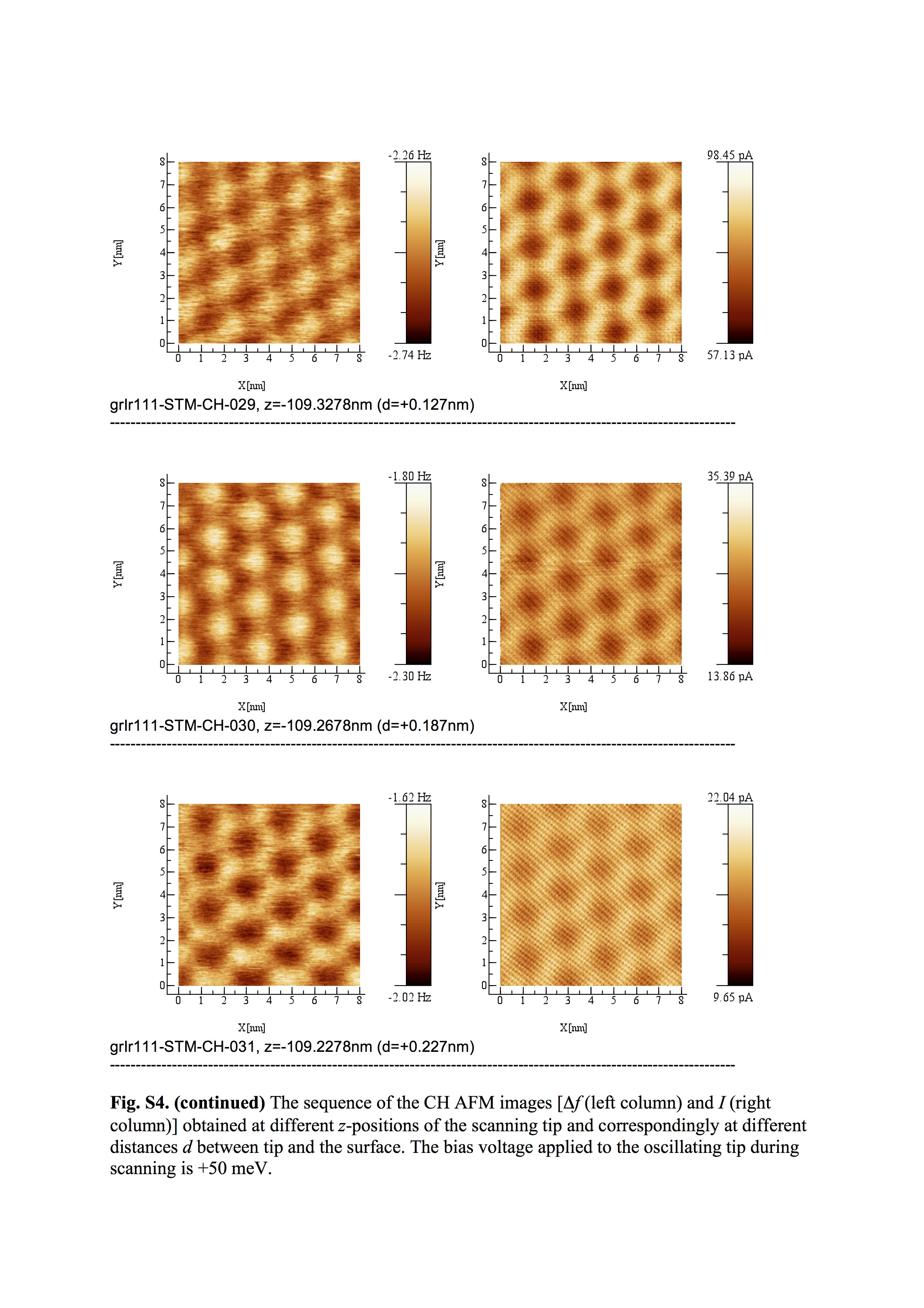}
\end{figure}

\end{document}